\documentclass[a4paper,12pt]{article}
\usepackage{graphicx}
\usepackage{amsmath}
\usepackage{amsthm}
\usepackage{amsfonts}
\usepackage{amssymb}
\usepackage{fancybox}
\usepackage{mathtools}
\usepackage{amsthm}
\usepackage{color}
\usepackage{natbib}
\usepackage{lscape}

\newtheorem{prop}{Proposition}

\newtheorem{lem}{Lemma}

\def\p#1{\mbox{${\rm P}\left[#1\right]$}}
\def\e#1{\mbox{${\rm E}\left[#1\right]$}}

\def\vec#1{\mbox{\boldmath$#1$}}

\numberwithin{equation}{section}
\begin{document}

\title{
On a class of binary regression models and their robust estimation
}
\author{
Kenichi Hayashi
\thanks{
Department of Mathematics, Keio University, 3-14-1 Hiyoshi, Kohoku-ku, Yokohama, 223-8522, Japan}\and
Shinto Eguchi
\thanks{
The Institute of Statistical Mathematics, 10-3 Midori-cho, Tachikawa, Tokyo 190-8562, Japan}
}
\date{}

\maketitle

\begin{abstract}
A robust estimation framework for binary regression models is studied, aiming to extend traditional approaches like logistic regression models.  
While previous studies largely focused on logistic models, we explore a broader class of models defined by general link functions. 
We incorporate various loss functions to improve estimation under model misspecification. 
Our investigation addresses robustness against outliers and model misspecifications, leveraging divergence-based techniques such as the $\beta$-divergence and $\gamma$-divergence, which generalize the maximum likelihood approach. 
These divergences introduce loss functions that mitigate the influence of atypical data points while retaining Fisher consistency.

We establish a theoretical property of the estimators under both correctly specified and misspecified models, analyzing their robustness through quantifying the effect of outliers in linear predictor. Furthermore, we uncover novel relationships between existing estimators and robust loss functions, identifying previously unexplored classes of robust estimators. 
Numerical experiments illustrate the efficacy of the proposed methods across various contamination scenarios, demonstrating their potential to enhance reliability in binary classification tasks. 
By providing a unified framework, this study highlights the versatility and robustness of divergence-based methods, offering insights into their practical application and theoretical underpinnings. 
\end{abstract}
\section{Introduction}
Regression models with categorical responses are statistical methods used in many fields, including medicine and economics and so forth. 
When the number of categories is two (dichotomous response), the logistic regression model is the most popular choice in practice. 
These methods are used to predict the response itself or the conditional probability of occurring a certain response given features. 
In recent years, with the development of machine learning, there are more flexible methods using kernel methods, random forest and so on.  
However, the logistic regression model is fundamental, often serving as the basis for these methods and their form also becomes more important as the calibration function for a learner's output. 

In this study, we provide a framework of robust estimation with a general binary linear regression model. 	
Although similar topics have been investigated from various viewpoints, most of them mainly focused on the logistic regression models. 
\cite{Bianco1996} proposed a robust estimator by bounding the log-likelihood to robustify for the logistic linear regression models and then implemented by \cite{Croux2003}. 
\cite{Victoria2002} investigated estimating equations weighted by the residuals. 
These methods require introduction of some bias correction terms to ensure Fisher consistency, which guarantees validity of estimation. 
The link function-based approaches also were proposed. 
\cite{Liu2004} and \cite{Ding2010} consider a family of t-distribution as a link function so as to robustify against outlying observations which features are far away from its concentrated region thanks to its heavy-tailed property. 
One of the most successful approach for robust regression is divergence-based methodologies. 
\cite{Ghosh2016} discussed a framework of the minimum density power divergence, that is proposed by \cite{Basu1998} in the framework of generalized linear models and analyzed its robustness by the influence function. 
The $L_2$-distance based estimation method is also included as a special case of this divergence \citep{Bondell2005,Chi2014}.
From another context, \cite{Hung2018} proposed a gamma-divergence based logistic regression model for mislabeled data, which contains a susceptible label in a dataset. 
Regression models for polytomous response are less investigated rather than one for binary response discussed above, several approaches have been studied recently \citep{Ghosh2016}.

While many robust methods for these models have been proposed in the past, our framework comprehends the vast range of the model family. 
We show that an interesting relation between the existing estimator and a robust loss function that has not been considered in the past. 
Furthermore, our analysis showed that a certain class of loss functions, which were never mentioned in previous studies, is also robust. 
Numerical experiments also confirm the behavior of robust loss functions for various misspecified model scenarios.

\section{Parameter estimation based on robust divergences}
\subsection{Problem setting and estimators}
Let $X$ be a $d$-dimensional feature vector and $Y$  be a  class label. 
Assume that $\mathcal{D}_n=\{(X_i,Y_i); i=1,\ldots,n\}$ is an I.I.D. sample of size $n$, that is, $(X_i,Y_i)$'s are independent copies of the random variable $(X,Y)$ from a probability distribution $p$ defined on $\mathcal{X}\times\mathcal{Y}$, where $\mathcal{X}\subset\mathbf{R}^d$ and $\mathcal{Y}=\{0,1\}$. 

The statistical issue that we focus in this paper is inference of the conditional probability $p(y|x)=\p{Y=y|X=x}$ via a statistical model
\begin{align}\label{eq:genbin}
q_G(y|x;\theta)=G(\theta^\top \tilde{x})^y\bar{G}(\theta^\top \tilde{x})^{1-y}
	\ \ {\rm with\ }\tilde{x}=(1,x^\top)^\top,
\end{align}
where $\theta$ is a $(d+1)$-dimensional parameter vector, $G$ is the cumulative distribution function of an arbitrary continuous random variable which support is $\mathbf{R}$, and $\bar{G}(z)=1-G(z)$ for any $z\in\mathbf{R}$. 
This model includes the logistic, probit, complementary log-log, regression models \citep{1989McCullaugh}.

For brevity, we write $q_G(y|x;\theta)$ as $q(y|x;\theta)$ or $q(y|x)$ unless it causes confusion.
In this study, we consider inference problems on $p(y|x)=\p{Y=y|X=x}$ using a parametric model $\{q(y|x;\theta); \theta\in\Theta\}$, where $\Theta$ is a parameter space.  
Given $G$, we estimate $\theta$ by minimizing an appropriate loss function $\Psi(y,\theta^\top\tilde{x})$. 
One of the most popular method is the maximum likelihood (ML) method. 
The ML estimator can be represented by minimizing sample mean of the the negative log-loss, which is given by 
\begin{align}\label{eq:neglogloss}
\Psi_{\rm ML}(y,\theta^\top\tilde{x})
=
-\log q(y|x;\theta)
=
-y\log G(\theta^\top \tilde{x}) -(1-y)\log \bar{G}(\theta^\top \tilde{x}).
\end{align}
The risk function by the loss \eqref{eq:neglogloss}, 
$\hat{R}_{\rm ML}(\theta;\mathcal{D}_n)=\displaystyle\frac{1}{n}\sum_{i=1}^n\Psi_{\rm ML}(Y_i,\theta^\top\tilde{X}_i)$, 
is proportional to the negative log-likelihood function, where $\tilde{X}_i$ is defined the same manner as $\tilde{x}$ in \eqref{eq:genbin}. 
To find the estimator of $\theta$ by the loss function \eqref{eq:neglogloss}, we typically solve the so-called likelihood equation
\begin{align}\label{eq:likeq}
\hat{B}_{\rm ML}(\theta;\mathcal{D}_n)
=
\nabla_\theta \hat{R}_{\rm ML}(\theta;\mathcal{D}_n)
=
-\sum_{i=1}^n S(Y_i,X_i;\theta)
=
0,
\end{align}
where $\nabla_\theta$ is the operator of partial derivative with respect to $\theta$ and $S(y,x;\theta)=\nabla_\theta \log q(y|x;\theta)$. 
Remark that the score function $S$ can be represented with $g$, the density function of $G$ by 
\begin{align}\label{eq:score.function}
S(y,x;\theta)
=
\left(y\frac{g(\theta^\top\tilde{x})}{G(\theta^\top\tilde{x})}
 -(1-y)\frac{g(\theta^\top\tilde{x})}{\bar{G}(\theta^\top\tilde{x})}\right)\tilde{x}.
\end{align}

When we expect the existence of outliers, the robust divergence can be utilized. 
In particular, the $\beta$-divergence or density power divergence \citep{Basu1998} is a popular choice. 
The $\beta$-divergence induced loss function for the binomial regression model is given as
\begin{align}\label{eq:beta}
\Psi_\beta(y,\theta^\top\tilde{x})
=
-\frac{1}{\beta} q(y|x;\theta)^{\beta}
 +\frac{1}{\beta+1}\sum_{m=0}^1 q(m|x;\theta)^{\beta+1}\ \ {\rm for}\ \beta>0.
\end{align}
\cite{Hung2018} established a $\gamma$-divergence based logistic linear regression models for mislabeled data, which is  one of the outlier mechanisms appearing in the binary data \citep{Hayashi2012}. 
The $\gamma$-divergence, proposed by \cite{Fujisawa2008}, induces a loss function 
\begin{align}\label{eq:gamma}
\Psi_\gamma(y,\theta^\top\tilde{x})
=
-\frac{1}{\gamma} q_\gamma(y|x;\theta)^{\frac{\gamma}{\gamma+1}},\ \ 
	{\rm where}\ \ 
q_\gamma(y|x;\theta)
=
\frac{q(y|x;\theta)^{\gamma+1}}
{\sum_{m=0}^1q(m|x;\theta)^{\gamma+1}}\ \ {\rm for}\ \gamma>0.
\end{align}
Note that the loss functions \eqref{eq:beta} and \eqref{eq:gamma} reduce to the negative log-likelihood function \eqref{eq:neglogloss} as $\beta\to0$ and $\gamma\to0$ respectively.
\subsection{Robustness of estimators} 
To analyze how different loss functions affect estimation, we examine the conditional expected score functions.
\begin{align}\label{eq:ml.score}
&B_{\rm ML}(\theta,x) 
= 
\nabla_\theta \e{\Psi_{\rm ML}(Y,X;\theta)|X=x}
=
\sum_{y=0}^1p(y|x)S(y,x;\theta).
\end{align}
Similarly, we define
\begin{align}\label{eq:beta.score}
&B_\beta(\theta,x) 
= 
\nabla_\theta \e{\Psi_\beta(Y,X;\theta)|X=x}
=
\sum_{y=0}^1\left\{q(y|x;\theta)-p(y|x)\right\}q(y|x;\theta)^{\beta}S(y,x;\theta),\ \ {\rm and}
\\
\label{eq:gamma.score}
&B_\gamma(\theta,x) 
= 
\nabla_\theta \e{\Psi_\gamma(Y,X;\theta)|X=x}
= 
-\frac{1}{\gamma+1}\sum_{y=0}^1
	p(y|x)q_\gamma(y|x;\theta)^{\frac{\gamma}{\gamma+1}}S_\gamma(y,x;\theta),
\end{align}
where $S_{\gamma}(y,x;\theta)=\nabla_\theta \log q_{\gamma}(y|x;\theta)$.
Consider the case of correctly specified model. 
That is, assume that there exists $\theta_0\in\Theta$ such that $p(x,y)=p(x)q(y| x;\theta_0)$ for any $(x,y)\in\mathcal{X}\times\mathcal{Y}$. 
Then, for example, it is easily seen that $B_{\rm ML}(\theta,x)=0$ for any $x\in\mathcal{X}$. 
Thus, the property called Fisher consistency holds: 
\begin{align}\label{eq:fisherconsistency}
\e{B_{\rm ML}(\theta_0,X)}=\e{B_\beta(\theta_0,X)}=\e{B_\gamma(\theta_0,X)}=0.
\end{align}
This justifies solving the score functions such as \eqref{eq:likeq}, the empirical version of $\e{B_{\rm ML}(\theta;X)}=0$. 

Next, consider the case where the model is misspecified (the case of such $\theta_0\in\Theta$ does not exist).
Since the property \eqref{eq:fisherconsistency} does not hold in this case, $\theta_\square^*$ is to be found such that $\e{B_\square(\theta_\square^*,X)}=0$, where ML, $\beta$ or $\gamma$ is substituted to $\square$ .
Hence, unboundedness of $\e{B_\square(\theta,X)}$ implies that the estimator of $\theta_\square^*$ can  be unstable. 
The loss function $\Psi_\square$ is a function of $y$ and $z=\theta^\top\tilde{x}$, and the behavior of $z$ is essential for unboundedness since $y$ is binary. 
Therefore, we quantify the contamination with respect to $z$ as follows.
For any loss function $\Psi_\square(y,z)$, we quantify the effect of contamination $z^{\prime}$ to $z=\theta^\top\tilde{x}$ by 
\begin{align}
\label{eq:IF.linear.pred}
b_\square(y,z,z')=
\lim_{\varepsilon\to0}
 \frac{\Psi_\square(y,(1-\varepsilon)z+\varepsilon z^{\prime})
 	-\Psi_\square(y,z)}{\varepsilon}
=
\psi_\square(y,z)(z^\prime-z),
\end{align}
where $\psi_\square(y,x;\theta)\tilde{x}=\nabla_\theta \Psi_\square(y,x;\theta)$, which can be found in Equations \eqref{eq:ml.score}--\eqref{eq:gamma.score}. 
While the vector $B_\square(\theta,x)$ represents the conditionally expected score function given $x$ and provides insights into the overall behavior of the model, the scalar $b_\square(y,z,z')$ offers a more comprehensive understanding of the sensitivity of the model's loss function to perturbation in the linear predictor $z$. 
This behavior of $b_\square$ is critical when evaluating the robustness of the model, especially under the misspecified scenarios. 
Additionally, by projecting the expected score functions onto the vector $\theta$, we have
\begin{align*}
\left|\e{B_\square(\theta,X)|X=x}^\top\theta\right|
=
\big|\e{b_\square(Y,z,0)|X=x}\big|.
\end{align*}
The left-hand side of the above equation represents the projection of the expected score function onto the vector $\theta$, providing a scalar measure of the score's behavior. 
By focusing on the equivalent quantity $b_\square(Y,z,0)$ on the right-hand side, we simplify the analysis of robustness and boundedness properties, which are essential for understanding the model's sensitivity to contamination in the linear predictor $z$. 
Moreover, the quantity \eqref{eq:IF.linear.pred} corresponds to the inner product of the main part (vector) of the influence function (IF) \citep{Hampel2011} and $\theta$. 
IF is $\psi_\square$ multiplied by a positive definite matrix from left. 
Also, since IF is a vector-valued function, evaluating unboundedness with a scalar value will clear up the discussion.
Obviously, if IF is bounded, then the limit \eqref{eq:IF.linear.pred} is also bounded (but not vice versa). 

\begin{figure}[!h]
\begin{center}
\begin{tabular}{cc}
\begin{minipage}{0.5\hsize}
\begin{center}
\includegraphics[width=7cm,height=7cm]{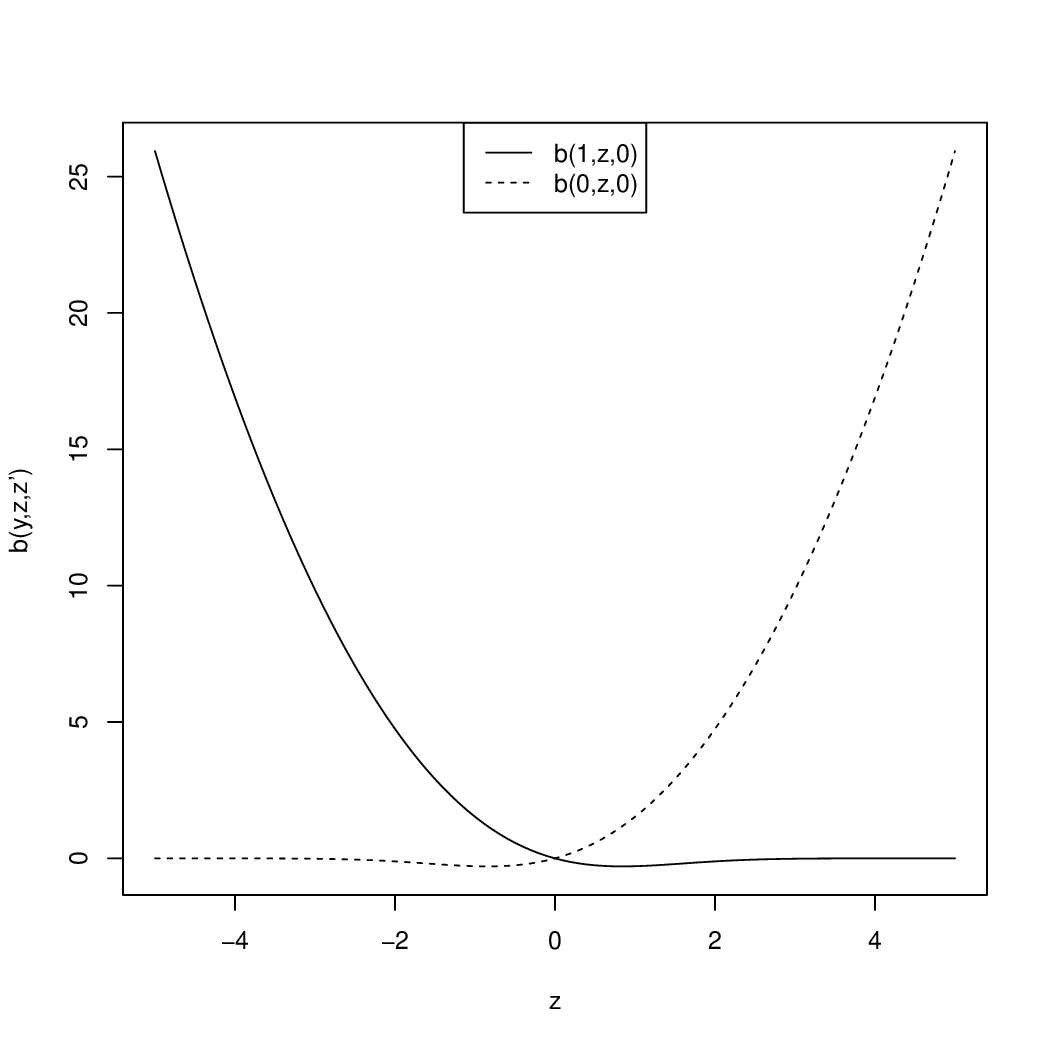}
\end{center}
\caption{Graph of $b_{\rm ML}$ when $G$ is the standard normal distribution (probit regression model).}
\label{fig:bml.probit}
\end{minipage}

\begin{minipage}{0.5\hsize}
\begin{center}
\includegraphics[width=7cm,height=7cm]{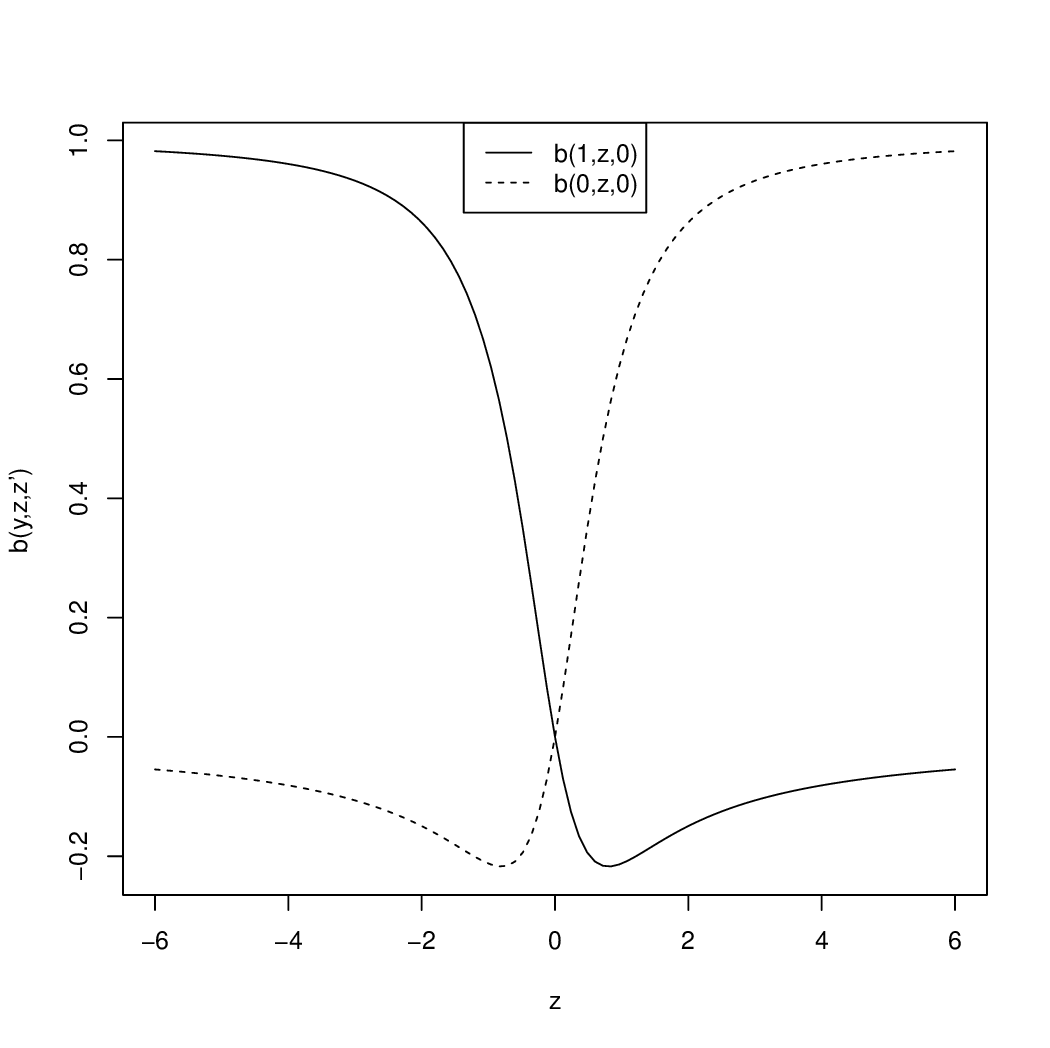}
\end{center}
\caption{Graph of $b_{\rm ML}$ when $G$ is the standard Cauchy distribution.}
\label{fig:bml.cauchit}
\end{minipage}
\end{tabular}
\end{center}
\end{figure}

Hence, we shall evaluate boundedness of the function $b_\square$.
The boundedness of $b_\square$ serves as a measure of the robustness of the model to outliers. 
Through understanding the behavior of $b_\square$, we can directly the stability of the robustness and its corresponding estimator without addressing $\e{B_\square(\theta,X)|X=x}$, which is difficult to analyze due to multidimensionality.  
Remark that the representation of the score \eqref{eq:score.function}, then we have
\begin{align*}
b_{\rm ML}(y,z,z')
=
\left(-y\frac{g(z)}{G(z)}
 +(1-y)\frac{g(z)}{\bar{G}(z)}\right)(z-z').
\end{align*}
Figure \ref{fig:bml.probit} depicts the behavior of $b_{\rm ML}(1,z)$  for a fixed $z'=0$ and $G(z)=\int_{-\infty}^z(2\pi)^{-1}\exp(-t^2/2)dt$
\footnote{
While it is important to evaluate the behaviors for larger value of $|z'|$, it is not essential for discussion of unboundedness of $b_{\rm ML}$.
} 
(probit regression models). 
The values of $b_{\rm ML}$ diverges negatively as $|z|\to\infty$, implying the lack of robustness of the ML method.

\begin{figure}[h]
\begin{center}
\begin{tabular}{cc}
\begin{minipage}{0.5\hsize}
\begin{center}
\includegraphics[width=7cm,height=7cm]{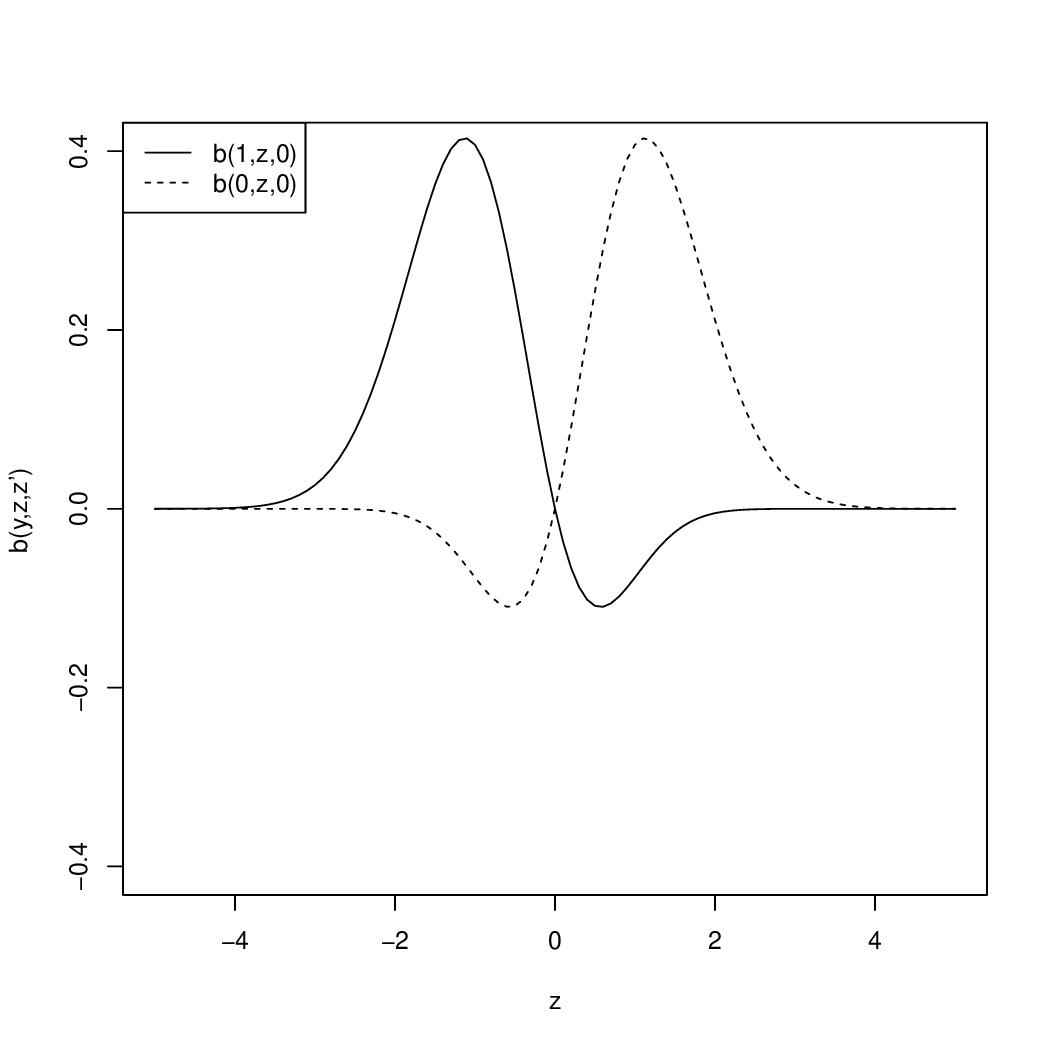}
\end{center}
\caption{Graph of $b_{\beta}$ when $G$ is the standard normal distribution ($\beta=1$).}
\label{fig:bml.probit_beta}
\end{minipage}

\begin{minipage}{0.5\hsize}
\begin{center}
\includegraphics[width=7cm,height=7cm]{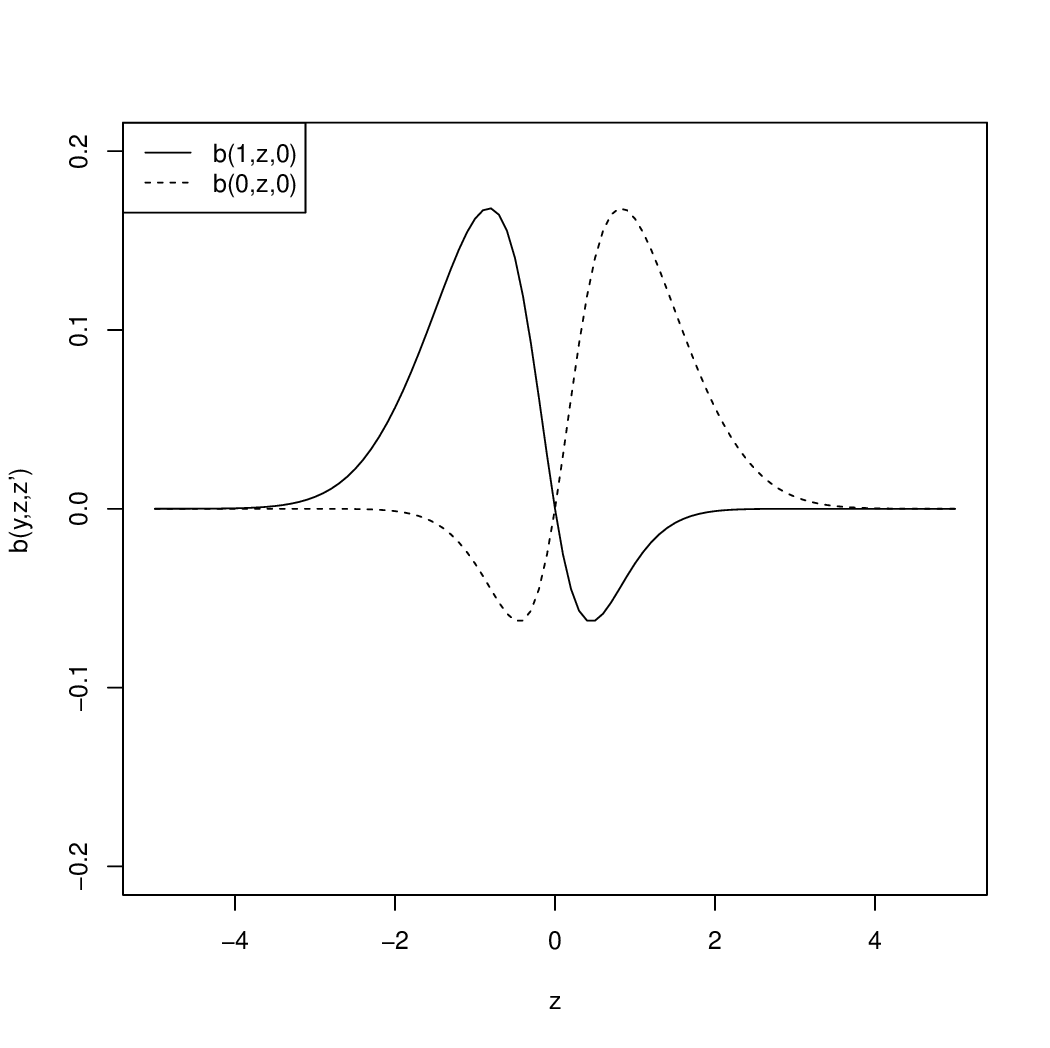}
\end{center}
\caption{Graph of $b_{\gamma}$ when $G$ is the standard normal distribution ($\gamma=1$).}
\label{fig:bml.probit_gamma}
\end{minipage}
\end{tabular}
\end{center}
\end{figure}
Here, we prepare the following lemma.
\begin{lem}\label{lem:dist.boundedness}
Define 
$L_1(c)
=
\lim_{z\to\infty}\displaystyle\frac{|z|g(z)}{\bar{G}(z)^c}
$ and
$L_2(c)
=
\lim_{z\to\infty}\displaystyle\frac{|z|g(z)}{G(-z)^c}
$,  
and assume that there exists $t>0$ such that
$
\lim_{z\to\infty}e^{tz}g(\pm z)=0.
$ 
Then, $L_1(c)=L_2(c)=0$ for $c\in(0,1)$ and $L_2(1)=L_2(1)=+\infty$.
\end{lem}
\noindent
The proof of Lemma \ref{lem:dist.boundedness} is provided in Appendix. 
The assumption in the lemma provides the sufficient condition that the $G$ is not a heavy-tailed distribution. 
The lemma indicates that $b_{\rm ML}$ is unbounded when $G(-z)$ and $\bar{G}(z)$ converge to 0 as $z\to\infty$ slower than an exponential order. 
In fact, as seen in Figure \ref{fig:bml.cauchit}, $b_{\rm ML}$ is bounded for when $G(z)=\pi^{-1}\tan^{-1}(z)$ (the model specified by \texttt{cauchit} for \texttt{glm} function in R). 
The cauchy distribution (t-distribution of the degree of freedom 1) is known as a heavy-tailed distribution and is investigated as a robust binary regression model \citep{Liu2004}. 

Let us return to the case where $b_{\rm ML}$ is unbounded as Figure \ref{fig:bml.probit}. 
The following proposition establishes the robustness of $\beta$- and $\gamma$-divergence based methods.
\begin{prop}\label{prop:bias.boundedness}
Let $G$ be a c.d.f satisfying the assumption in Lemma \ref{lem:dist.boundedness} and $z'$ be a fixed scalar. 
Then, $|b_\beta(y,\theta^\top\tilde{x},z')|$ is bounded for any $(x,y)\in\mathcal{X}\times\mathcal{Y}$ when $\beta>0$. 
Similarly, $|b_\gamma(y,\theta^\top\tilde{x},z')|$ is bounded when $\gamma>0$ or $\gamma<-1$.
\end{prop}
\noindent
The proof of Proposition \ref{prop:bias.boundedness} is provided in Appendix. 
As shown in Figures \ref{fig:bml.probit_beta} and \ref{fig:bml.probit_gamma}, $b_\beta$ or $b_\gamma$ is unbounded even when $b_{\rm ML}$ is unbounded. 
Remark that the fact of Proposition \ref{prop:bias.boundedness} is not a novel for the $\beta$-loss function, and nor for the $\gamma$-loss function for $\gamma>0$. 
It is rather more interesting that such robustness holds for negative value of $\gamma$. 
We observe that in the $\beta$-divergence, the exponent $\beta$ directly modifies the loss function $\Psi_\beta(y,\theta^\top\tilde{x})$, controlling how strongly discrepancies between the model and data contribute to the estimation. 
In contrast, in the $\gamma$-divergence, the parameter $\gamma$ appears in the ratio term $\displaystyle\frac{\gamma}{\gamma+1}$ within $\Psi_\beta(y,\theta^\top\tilde{x})$. 
This ratio remains positive when $\gamma>0$ or $\gamma<-1$, ensuring that the loss function remains well-behaved under contamination. 
Consequently, when $\gamma<-1$, the contribution of extreme outliers is naturally downweighted, enhancing robustness. 
This observation provides additional intuition for why $\gamma$-divergence estimators exhibit robustness in the region $\gamma<-1$.
For example, when we formally substitute $\gamma=-2$ to the loss function \eqref{eq:gamma}, we have

\begin{align}\label{eq:harmonic}
\Psi_{\gamma=-2}(y,\theta^\top\tilde{x})
&=
\left(\frac{1}{q(y|x;\theta)^2}\right)
	\left(\frac{1}{q(y|x;\theta)}+\frac{1}{q(1-y|x;\theta)}\right)^{-2}
\end{align}
This loss function is derived by the cross-entropy
\begin{align*}
H_{\rm HM}(q(\cdot|x;\theta),p(\cdot|x))
=
\left(\sum_{y=0}^{1}\frac{p(y|x)}{q(y|x;\theta)^2}\right)
	\left(\sum_{m=0}^{1}\frac{1}{q(m|x;\theta)}\right)^{-2}.
\end{align*}
The suffix ``HM'' comes from the fact that $H_{\rm HM}(p(\cdot|x),p(\cdot|x))$ is proportional to harmonic mean of $p(1|x)$ and $p(0|x)$. 
Note that $H_{\rm HM}(p(\cdot|X),p(\cdot|X))$ is maximized when $p(y|x)$ is the uniform distribution, $p(1|x)=p(0|x)=\dfrac12$ for any $x\in\mathcal{X}$. 
It is well known that the classical entropy has the same property \citep{MacKay2003}. 
From easy calculation, we also write the empirical risk function based on the loss \eqref{eq:harmonic} as 
\begin{align}\label{eq:harmonic.l2e}
\hat{R}_{\gamma=-2}(\theta;\mathcal{D}_n)
=
\frac{1}{n}\sum_{i=1}^n
\Psi_{\gamma=-2}(Y_i,\theta^\top\tilde{X}_i)
=
\frac{1}{n}\sum_{i=1}^n\left(Y_i-q(1|X_i;\theta)\right)^2.
\end{align}
The rightmost side in Equation \eqref{eq:harmonic.l2e} is the Brier score, which has been a well-known index in decision making \citep{Merkle2013}. 
Although the Brier score is used for evaluation of models $q$ rather than a risk function to be optimized, Chi and Scott (2014) proposed an estimator obtained by minimizing a ``distance'' that is essentially equivalent to the risk function \eqref{eq:harmonic.l2e}. 
The robustness of the loss function \eqref{eq:harmonic} is also understood by the equivalence of $\Psi_\beta$ when $\beta=1$. 
That is, $\Psi_{\beta=1}(Y,\theta^\top\tilde{X})=\left(Y-q(1|X;\theta)\right)^2-\dfrac{1}{2}$ 
and hence the behavior of $b_{\gamma=-2}$ is proportional to Figure \ref{fig:bml.probit_beta} (when $G$ is the standard normal distribution). 
This connection is interesting in the sense that, to the best of our knowledge, the case of nonnegative $\gamma$ was not studied in previous research. 
This would be because they begin with the regression models with continuous responses, where the sum $\sum_{m=0}^{1}q(m|x;\theta)^{\gamma+1}$ in the function \eqref{eq:gamma} is replaced with an integral over the marginal distribution of $Y$.
In this case, its integrability does not necessarily guaranteed for negative $\gamma$. 
On the other hand, when the response variables have a finite number of categories, the risk function \eqref{eq:harmonic.l2e} always exists with the assumption $q(y|x)>0$ for any $(x,y)\in\mathcal{X}\times\mathcal{Y}$. 

Another interesting point is the case
\begin{align}\label{eq:geometric}
\lim_{\gamma\to-1}\Psi_{\gamma}(y,\theta^\top\tilde{x})
&=
\frac12\left(\frac{q(1-y|x;\theta)}{q(y|x;\theta)}\right)^\frac{1}{2}, 
\end{align}
which is derived by the geometric mean induced cross-entropy
\begin{align*}
H_{\rm GM}(q(\cdot|x;\theta),p(\cdot|x))
=
\frac{1}{2}\left(\prod_{y=0}^{1}\frac{p(y|x)}{q(y|x;\theta)}\right)
	\left(\prod_{m=0}^{1}q(m|x;\theta)\right)^{\frac{1}{2}}.
\end{align*}
As Proposition \ref{prop:bias.boundedness}, however, the loss \eqref{eq:geometric} does not have robustness.

\section{Numerical experiments}
To examine the behavior of each method mentioned in this study, we conducted numerical experiments via a Monte Carlo simulation.  
All calculations reported in this section were performed using R version 4.1.4 \citep{R2014}.
The aim of the experiments is to confirm the robustness property can be seen in finite sample. 
We mainly consider two scenarios: one is that the model correctly specified but contaminated data are included (Scenario 1) and the other is misspecified data (Scenario 2). 
\subsection{Settings}
For Scenario 1, we first generate $n$ independent copies of $X\sim$Unif$[-3,3]^2$ as a feature vector. 
Then $Y$ is generated by $P[Y=1|X]=\dfrac{e^{\theta_*^\top\tilde{X}}}{1+e^{\theta_*^\top\tilde{X}}}$, 
where $\theta_*=(0,a,-a)^\top$ for several $a>0$. 
$n_{\rm out}$ is a randomly generated by the binomial distribution ${\rm Bin}(n,p_{\rm out})$ and $n_{\rm out}$ outcomes $Y$ are replaced by $Y_{\rm out}$ generated by the probability $P[Y_{\rm out}=1|X]=\dfrac{e^{-\theta_*^\top\tilde{X}}}{1+e^{-\theta_*^\top\tilde{X}}}$. 

For Scenario 2, we consider two cases A and B. 
Case A is generated from a binormal (two normal populations) model . 
$n_1$ is generated by the binomial distribution ${\rm Bin}(n,r)$. 
Then $n_1$ feature vectors are generated by $N(\vec{\mu}_1,I_2)$ where $I_2$ is the two-dimensional unit matrix. 
Similarly, $n_0=n-n_1$ feature vectors are generated by $N(\vec{\mu}_0,sI_2)$ for $s>0$. 
Here, we specify that two mean vectors are specified by their distance $D=\|\vec{\mu}_1-\vec{\mu}_0\|$. 
Case B is generated from two t-distributions. 
$n_1$ and $n_0$ follows the same as Case A, and feature vectors of $Y=1$ and $Y=0$ follows t-distribution with the degrees of freedom $\nu_1$ and $\nu_0$, respectively. 
\subsection{Results}
For all Scenarios and Cases, we obtained the estimates by generating training data 1,000 times for each setting with $n=400$. 
The results were evaluated by assessing the correct classification rate on test data (sample size 50,000).

In Scenario 1, maximum likelihood estimation performs worse in the presence of outliers, suggesting that robust divergence-based methods provide better accuracy under contamination (Table \ref{tab:scenario1}). 
In particular, it is shown that the larger the ratio of outliers $p_{\rm out}$, the better results are given when $\beta$ or $|\gamma|$ is large. 
The best results are obtained when $\gamma=-2$ in some settings, and these values are almost the same as those obtained when $\beta=1$,  as described in the previous section. 
It is therefore worth noting that the negative $\gamma$ (especially when $-2$) can give useful results.

In Scenario 2 Case A, the model is correctly specified when $s=1$, so it is natural that the likelihood-based method gives the best results in this case (see Table \ref{tab:scenario2_casea}). 
When $s\neq1$, i.e., when the linear model is not the optimal model, it is suggested that the robust divergence-based method may give better results. 
In particular, when the overlap between the two groups is large (i.e., when the classification rate is on average low), these methods have an advantage over the method based on log-likelihood.
In Scenario 2 Case B, there were many cases where the algorithm did not work correctly because extremely large values of covariates were generated when the degrees of freedom of the t-distribution were small.  
If such cases exceeded 100 out of 1,000 times, comparison of averages would be inappropriate, so they are indicated by ``--''.  
In the case of two bivariate t-distributions, when the degrees of freedom of the two groups are equal, the linear model is correctly specified. 
Hence the estimation of the linear model based on the log-likelihood tends to give the best results when $\nu_1=\nu_2=20$. 
However, the algorithm does not work correctly in many cases. 
On the other hand, methods based on $\beta$-divergence tend to achieve stable estimation and good classification rates when the class probabilities are balanced. 
Similarly, it can be seen that estimation with $\gamma>0$ gives good results when the model is misspecified. 
On the other hand, when $\gamma$ is negative, the algorithm often does not work as well as methods based on log-likelihood.

\begin{table}[htbp]
\begin{center}
\scriptsize{
\begin{tabular}{cc||c|ccc|ccccccc} 
 &  & \multicolumn{1}{c|}{MLE} & \multicolumn{3}{c|}{$\beta$} & \multicolumn{7}{c}{$\gamma$} \\
$a$ & $p_{\rm out}$ & \multicolumn{1}{c|}{} & 0.25 & 0.50 & 1.00 & -2.00 & -1.50 & -1.00 & -0.50 & 0.25 & 0.50 & 1.00 \\\hline\hline
\multicolumn{1}{c}{} & 0.05 & 81.218 & 81.227 & \bf{81.230} & 81.228 & 81.228 & 81.222 & 81.173 & 81.173 & 81.228 & 81.228 & 81.211 \\
\multicolumn{1}{c}{1/3} & 0.10 & 81.143 & 81.163 & 81.174 & 81.179 & \bf{81.179} & 81.150 & 81.088 & 81.088 & 81.165 & 81.177 & 81.175 \\
\multicolumn{1}{c}{} & 0.15 & 81.043 & 81.069 & 81.085 & 81.100 & 81.100 & 81.048 & 80.985 & 80.985 & 81.072 & 81.093 & \bf{81.107} \\
\multicolumn{1}{c}{} & 0.20 & 80.947 & 80.976 & 80.995 & 81.017 & 81.017 & 80.951 & 80.887 & 80.887 & 80.980 & 81.008 & \bf{81.039} \\\hline
\multicolumn{1}{c}{} & 0.05 & 81.219 & 81.230 & \bf{81.233} & 81.231 & 81.231 & 81.224 & 81.175 & 81.175 & 81.230 & 81.232 & 81.214 \\
\multicolumn{1}{c}{1/2} & 0.10 & 81.111 & 81.130 & 81.140 & \bf{81.148} & 81.148 & 81.116 & 81.057 & 81.057 & 81.132 & 81.144 & 81.148 \\
\multicolumn{1}{c}{} & 0.15 & 80.982 & 81.007 & 81.023 & 81.038 & 81.038 & 80.987 & 80.924 & 80.924 & 81.011 & 81.032 & \bf{81.053} \\
\multicolumn{1}{c}{} & 0.20 & 80.820 & 80.853 & 80.877 & 80.906 & 80.906 & 80.824 & 80.760 & 80.760 & 80.858 & 80.895 & \bf{80.945} \\\hline
\multicolumn{1}{c}{} & 0.05 & 81.194 & 81.204 & \bf{81.206} & 81.205 & 81.205 & 81.199 & 81.147 & 81.147 & 81.204 & 81.205 & 81.189 \\
\multicolumn{1}{c}{1} & 0.10 & 81.083 & 81.102 & 81.112 & \bf{81.121} & 81.121 & 81.088 & 81.034 & 81.034 & 81.104 & 81.116 & 81.119 \\
\multicolumn{1}{c}{} & 0.15 & 80.937 & 80.962 & 80.979 & 80.996 & 80.996 & 80.941 & 80.884 & 80.884 & 80.966 & 80.988 & \bf{81.010} \\
\multicolumn{1}{c}{} & 0.20 & 80.719 & 80.746 & 80.765 & 80.790 & 80.790 & 80.722 & 80.668 & 80.668 & 80.750 & 80.781 & \bf{80.827} \\\hline
\multicolumn{1}{c}{} & 0.05 & 81.179 & 81.192 & 81.198 & \bf{81.199} & 81.199 & 81.184 & 81.135 & 81.135 & 81.192 & 81.199 & 81.188 \\
\multicolumn{1}{c}{3/2} & 0.10 & 81.069 & 81.082 & 81.090 & 81.095 & \bf{81.095} & 81.073 & 81.028 & 81.028 & 81.084 & 81.092 & 81.090 \\
\multicolumn{1}{c}{} & 0.15 & 80.901 & 80.922 & 80.935 & 80.947 & 80.947 & 80.905 & 80.861 & 80.861 & 80.925 & 80.942 & \bf{80.958} \\
\multicolumn{1}{c}{} & 0.20 & 80.648 & 80.672 & 80.689 & 80.709 & 80.709 & 80.651 & 80.603 & 80.603 & 80.677 & 80.703 & \bf{80.743} \\\hline
\multicolumn{1}{c}{} & 0.05 & 81.185 & 81.194 & \bf{81.196} & 81.194 & 81.194 & 81.188 & 81.147 & 81.147 & 81.195 & 81.195 & 81.179 \\
\multicolumn{1}{c}{2} & 0.10 & 81.086 & 81.102 & 81.111 & 81.119 & \bf{81.119} & 81.090 & 81.038 & 81.038 & 81.104 & 81.114 & 81.114 \\
\multicolumn{1}{c}{} & 0.15 & 80.899 & 80.917 & 80.929 & 80.940 & 80.940 & 80.902 & 80.856 & 80.856 & 80.920 & 80.935 & \bf{80.945} \\
\multicolumn{1}{c}{} & 0.20 & 80.645 & 80.664 & 80.677 & 80.690 & 80.690 & 80.646 & 80.607 & 80.607 & 80.667 & 80.686 & \bf{80.705} \\

\end{tabular}
}
\end{center}
\caption{
Average accuracy for each estimation method in Scenario 1 with $n=400$ (\%). 
ML indicates the result based on the likelihood. 
The bold is the largest value for each setting. }
\label{tab:scenario1}
\end{table}

\begin{table}[htbp]
\begin{center}
\scriptsize{
\begin{tabular}{ccc||c|ccc|ccccccc} 
 &  &  & \multicolumn{1}{c|}{ML} & \multicolumn{3}{c|}{$\beta$} & \multicolumn{7}{c}{$\gamma$} \\
$r$ & $D$ & $s$ & \multicolumn{1}{c|}{} & 0.25 & 0.50 & 1.00 & -2.00 & -1.50 & -1.00 & -0.50 & 0.25 & 0.50 & 1.00 \\\hline\hline
\multicolumn{1}{c}{} & \multicolumn{1}{c}{} & 1/3 & \bf{98.546} & 98.527 & 98.499 & 98.466 & 98.489 & 98.537 & 98.445 & 98.445 & 98.526 & 98.493 & 98.448 \\
\multicolumn{1}{c}{} & \multicolumn{1}{c}{2} & 1 & \bf{95.979} & 95.972 & 95.958 & 95.934 & 95.939 & 95.976 & 95.924 & 95.924 & 95.972 & 95.952 & 95.893 \\
\multicolumn{1}{c}{} & \multicolumn{1}{c}{} & 3 & 90.538 & 90.523 & 90.509 & 90.491 & 90.491 & 90.529 & \bf{90.547} & \bf{90.547} & 90.522 & 90.503 & 90.464 \\
\multicolumn{1}{c}{} & \multicolumn{1}{c}{} & 1/3 & 99.789 & 99.753 & 99.741 & 99.738 & 99.774 & \bf{99.793} & 99.742 & 99.754 & 99.750 & 99.739 & 99.731 \\
\multicolumn{1}{c}{0.1} & \multicolumn{1}{c}{3} & 1 & \bf{98.967} & 98.949 & 98.929 & 98.899 & 98.922 & 98.962 & 98.914 & 98.915 & 98.946 & 98.922 & 98.870 \\
\multicolumn{1}{c}{} & \multicolumn{1}{c}{} & 3 & 95.603 & 95.616 & 95.614 & 95.602 & 95.603 & 95.610 & 95.507 & 95.507 & \bf{95.616} & 95.613 & 95.577 \\
\multicolumn{1}{c}{} & \multicolumn{1}{c}{} & 1/3 & 99.968 & 99.969 & 99.965 & 99.960 & 99.965 & 99.968 & \bf{99.980} & 99.977 & 99.968 & 99.964 & 99.956 \\
\multicolumn{1}{c}{} & \multicolumn{1}{c}{4} & 1 & \bf{99.790} & 99.718 & 99.718 & 99.720 & 99.770 & 99.783 & 99.752 & 99.752 & 99.715 & 99.716 & 99.724 \\
\multicolumn{1}{c}{} & \multicolumn{1}{c}{} & 3 & \bf{98.479} & 98.474 & 98.456 & 98.423 & 98.431 & 98.479 & 98.414 & 98.414 & 98.473 & 98.450 & 98.375 \\\hline
\multicolumn{1}{c}{} & \multicolumn{1}{c}{} & 1/3 & 96.386 & \bf{96.388} & 96.374 & 96.348 & 96.358 & 96.387 & 96.242 & 96.242 & 96.387 & 96.370 & 96.308 \\
\multicolumn{1}{c}{} & \multicolumn{1}{c}{2} & 1 & \bf{92.017} & 92.012 & 92.004 & 91.989 & 91.989 & 92.015 & 91.967 & 91.967 & 92.011 & 91.999 & 91.959 \\
\multicolumn{1}{c}{} & \multicolumn{1}{c}{} & 3 & 85.401 & 85.486 & 85.520 & 85.536 & \bf{85.536} & 85.444 & 84.933 & 84.933 & 85.491 & 85.528 & 85.531 \\
\multicolumn{1}{c}{} & \multicolumn{1}{c}{} & 1/3 & \bf{99.576} & 99.504 & 99.495 & 99.487 & 99.539 & 99.574 & 99.464 & 99.489 & 99.510 & 99.494 & 99.477 \\
\multicolumn{1}{c}{0.5} & \multicolumn{1}{c}{3} & 1 & 98.202 & 98.194 & 98.179 & 98.150 & 98.169 & \bf{98.203} & 98.141 & 98.141 & 98.194 & 98.172 & 98.109 \\
\multicolumn{1}{c}{} & \multicolumn{1}{c}{} & 3 & 94.105 & 94.116 & 94.110 & 94.092 & 94.093 & 94.112 & 93.895 & 93.895 & \bf{94.116} & 94.106 & 94.054 \\
\multicolumn{1}{c}{} & \multicolumn{1}{c}{} & 1/3 & 99.947 & 99.952 & 99.954 & 99.947 & 99.950 & 99.951 & \bf{99.964} & 99.962 & 99.949 & 99.951 & 99.944 \\
\multicolumn{1}{c}{} & \multicolumn{1}{c}{4} & 1 & \bf{99.697} & 99.611 & 99.601 & 99.591 & 99.672 & 99.691 & 99.629 & 99.628 & 99.608 & 99.593 & 99.582 \\
\multicolumn{1}{c}{} & \multicolumn{1}{c}{} & 3 & 98.020& \bf{98.029} & 98.009 & 97.975 & 97.974 & 98.017 & 97.930 & 97.930 & 98.029 & 98.004 & 97.932 \\\hline
\multicolumn{1}{c}{} & \multicolumn{1}{c}{} & 1/3 & 97.251 & 97.253 & 97.244 & 97.220 & 97.236 & \bf{97.255} & 97.172 & 97.172 & 97.253 & 97.241 & 97.191 \\
\multicolumn{1}{c}{} & \multicolumn{1}{c}{2} & 1 & \bf{95.984} & 95.978 & 95.967 & 95.944 & 95.947 & 95.982 & 95.935 & 95.935 & 95.978 & 95.961 & 95.908 \\
\multicolumn{1}{c}{} & \multicolumn{1}{c}{} & 3 & \bf{95.101} & 95.101 & 95.083 & 95.051 & 95.054 & 95.099 & 94.927 & 94.927 & 95.101 & 95.080 & 95.013 \\
\multicolumn{1}{c}{} & \multicolumn{1}{c}{} & 1/3 & \bf{99.651} & 99.586 & 99.574 & 99.562 & 99.611 & 99.644 & 99.575 & 99.589 & 99.581 & 99.570 & 99.548 \\
\multicolumn{1}{c}{0.9} & \multicolumn{1}{c}{3} & 1 & \bf{98.960} & 98.934 & 98.906 & 98.885 & 98.918 & 98.953 & 98.903 & 98.904 & 98.932 & 98.898 & 98.864 \\
\multicolumn{1}{c}{} & \multicolumn{1}{c}{} & 3 & \bf{97.751} & 97.740 & 97.712 & 97.680 & 97.695 & 97.739 & 97.628 & 97.628 & 97.739 & 97.708 & 97.651 \\
\multicolumn{1}{c}{} & \multicolumn{1}{c}{} & 1/3 & 99.960 & 99.957 & 99.954 & 99.952 & 99.959 & 99.959 & \bf{99.969} & 99.969 & 99.956 & 99.953 & 99.954 \\
\multicolumn{1}{c}{} & \multicolumn{1}{c}{4} & 1 & \bf{99.835} & 99.723 & 99.719 & 99.717 & 99.822 & 99.833 & 99.758 & 99.758 & 99.724 & 99.717 & 99.721 \\
\multicolumn{1}{c}{} & \multicolumn{1}{c}{} & 3 & \bf{99.154} & 99.126 & 99.102 & 99.084 & 99.129 & 99.149 & 99.060 & 99.060 & 99.126 & 99.100 & 99.069 \\
\end{tabular}
}
\end{center}
\caption{
Average accuracy for each estimation method in Scenario 2 Case A with $n=400$ (\%). 
ML indicates the result based on the likelihood. 
The bold is the largest value for each setting. }
\label{tab:scenario2_casea}
\end{table}

\begin{table}[htbp]
\begin{center}
\scriptsize{
\begin{tabular}{cccc||c|ccc|ccccccc} 
 &  &  &  & \multicolumn{1}{c|}{ML} & \multicolumn{3}{c|}{$\beta$} & \multicolumn{7}{c}{$\gamma$} \\
$r$ & $D$ & $\nu_1$ & $\nu_2$ & \multicolumn{1}{c|}{} & 0.25 & 0.50 & 1.00 & --2.00 & --1.50 & --1.00 & --0.50 & 0.25 & 0.50 & 1.00 \\\hline\hline
	\multicolumn{1}{c}{} & \multicolumn{1}{c}{} & 2 & 2  & -- & 89.380 & 89.552 & 89.690 & -- & -- & -- & -- & 89.392 & 89.599 & \bf{89.800} \\
	\multicolumn{1}{c}{} & \multicolumn{1}{c}{} & 2 & 7  & -- & 94.058 & 94.084 & 94.084 & -- & -- & 91.438 & 91.438 & 94.062 & \bf{94.088} & 94.062 \\
	\multicolumn{1}{c}{} & \multicolumn{1}{c}{2} & 2 & 20  & -- & 95.132 & \bf{95.133} & 95.114 & -- & -- & 92.412 & 92.412 & 95.133 & 95.128 & 95.080 \\
	\multicolumn{1}{c}{} & \multicolumn{1}{c}{} & 7 & 7  & 94.214 & 94.266 & 94.275 & 94.269 & 94.269 & 94.249 & 93.934 & 93.934 & 94.268 & \bf{94.277} & 94.245 \\
	\multicolumn{1}{c}{} & \multicolumn{1}{c}{} & 7 & 20  & 95.341 & 95.345 & 95.333 & 95.310 & 95.313 & \bf{95.345} & 95.178 & 95.178 & 95.344 & 95.329 & 95.274 \\
	\multicolumn{1}{c}{} & \multicolumn{1}{c}{} & 20 & 20  & 95.406 & 95.405 & 95.395 & 95.375 & 95.376 & \bf{95.407} & 95.327 & 95.327 & 95.405 & 95.390 & 95.340 \\
	\multicolumn{1}{c}{} & \multicolumn{1}{c}{} & 2 & 2  & -- & 93.341 & 93.681 & 93.789 & -- & -- & -- & -- & 93.371 & 93.719 & \bf{93.802} \\
	\multicolumn{1}{c}{} & \multicolumn{1}{c}{} & 2 & 7  & -- & 97.257 & 97.244 & 97.214 & -- & -- & 94.007 & 94.007 & \bf{97.257} & 97.240 & 97.183 \\
	\multicolumn{1}{c}{0.1} & \multicolumn{1}{c}{3} & 2 & 20  & -- & 97.930 & 97.912 & 97.886 & -- & -- & 94.912 & 94.912 & \bf{97.930} & 97.908 & 97.865 \\
	\multicolumn{1}{c}{} & \multicolumn{1}{c}{} & 7 & 7  & 97.713 & \bf{97.722} & 97.707 & 97.680 & 97.690 & 97.722 & 97.482 & 97.481 & 97.722 & 97.703 & 97.646 \\
	\multicolumn{1}{c}{} & \multicolumn{1}{c}{} & 7 & 20  & \bf{98.422} & 98.416 & 98.394 & 98.364 & 98.382 & 98.419 & 98.313 & 98.313 & 98.415 & 98.389 & 98.340 \\
	\multicolumn{1}{c}{} & \multicolumn{1}{c}{} & 20 & 20  & \bf{98.566} & 98.554 & 98.529 & 98.503 & 98.517 & 98.560 & 98.491 & 98.491 & 98.554 & 98.522 & 98.478 \\
	\multicolumn{1}{c}{} & \multicolumn{1}{c}{} & 2 & 2  & -- & 95.950 & 96.008 & 95.984 & -- & -- & -- & -- & 95.959 & \bf{96.018} & 95.940 \\
	\multicolumn{1}{c}{} & \multicolumn{1}{c}{} & 2 & 7  & -- & 98.584 & 98.561 & 98.538 & -- & -- & 95.321 & 95.323 & \bf{98.584} & 98.556 & 98.531 \\
	\multicolumn{1}{c}{} & \multicolumn{1}{c}{4} & 2 & 20  & -- & \bf{98.969} & 98.950 & 98.939 & -- & -- & 95.592 & 95.592 & 98.968 & 98.948 & 98.938 \\
	\multicolumn{1}{c}{} & \multicolumn{1}{c}{} & 7 & 7  & -- & \bf{99.069} & 99.038 & 99.024 & -- & -- & 98.983 & 98.983 & 99.068 & 99.032 & 99.015 \\
	\multicolumn{1}{c}{} & \multicolumn{1}{c}{} & 7 & 20  & -- & \bf{99.430} & 99.412 & 99.401 & -- & -- & 99.408 & 99.407 & 99.426 & 99.407 & 99.396 \\
	\multicolumn{1}{c}{} & \multicolumn{1}{c}{} & 20 & 20  & -- & 99.537 & 99.519 & 99.514 & -- & -- & \bf{99.540} & 99.537 & 99.535 & 99.520 & 99.511 \\\hline
	\multicolumn{1}{c}{} & \multicolumn{1}{c}{} & 2 & 2  & -- & 85.138 & 85.140 & 85.127 & -- & -- & -- & -- & 85.137 & \bf{85.141} & 85.108 \\
	\multicolumn{1}{c}{} & \multicolumn{1}{c}{} & 2 & 7  & -- & 87.473 & \bf{87.501} & 87.496 & -- & -- & 78.256 & 78.256 & 87.477 & 87.500 & 87.472 \\
	\multicolumn{1}{c}{} & \multicolumn{1}{c}{2} & 2 & 20  & -- & 88.122 & 88.169 & \bf{88.175} & -- & -- & 78.311 & 78.311 & 88.129 & 88.169 & 88.161 \\
	\multicolumn{1}{c}{} & \multicolumn{1}{c}{} & 7 & 7  & 89.839 & \bf{89.843} & 89.835 & 89.819 & 89.819 & 89.840 & 89.637 & 89.637 & 89.842 & 89.831 & 89.793 \\
	\multicolumn{1}{c}{} & \multicolumn{1}{c}{} & 7 & 20  & 90.540 & \bf{90.550} & 90.543 & 90.528 & 90.528 & 90.547 & 90.294 & 90.294 & 90.550 & 90.539 & 90.495 \\
	\multicolumn{1}{c}{} & \multicolumn{1}{c}{} & 20 & 20  & \bf{91.243} & 91.241 & 91.231 & 91.214 & 91.215 & 91.242 & 91.154 & 91.154 & 91.240 & 91.227 & 91.186 \\
	\multicolumn{1}{c}{} & \multicolumn{1}{c}{} & 2 & 2  & -- & \bf{91.424} & 91.415 & 91.391 & -- & -- & -- & -- & 91.424 & 91.413 & 91.361 \\
	\multicolumn{1}{c}{} & \multicolumn{1}{c}{} & 2 & 7  & -- & \bf{93.843} & 93.842 & 93.819 & -- & -- & 81.296 & 81.296 & 93.842 & 93.837 & 93.790 \\
	\multicolumn{1}{c}{0.5} & \multicolumn{1}{c}{3} & 2 & 20  & -- & 94.468 & \bf{94.473} & 94.461 & -- & -- & 80.932 & 80.932 & 94.472 & 94.472 & 94.437 \\
	\multicolumn{1}{c}{} & \multicolumn{1}{c}{} & 7 & 7  & \bf{96.304} & 96.296 & 96.278 & 96.245 & 96.250 & 96.296 & 96.087 & 96.087 & 96.296 & 96.272 & 96.205 \\
	\multicolumn{1}{c}{} & \multicolumn{1}{c}{} & 7 & 20  & 96.922 & \bf{96.926} & 96.908 & 96.880 & 96.892 & 96.925 & 96.629 & 96.629 & 96.926 & 96.904 & 96.847 \\
	\multicolumn{1}{c}{} & \multicolumn{1}{c}{} & 20 & 20  & \bf{97.564} & 97.560 & 97.542 & 97.508 & 97.520 & 97.564 & 97.463 & 97.463 & 97.559 & 97.536 & 97.471 \\
	\multicolumn{1}{c}{} & \multicolumn{1}{c}{} & 2 & 2  & -- & \bf{94.558} & 94.542 & 94.516 & -- & -- & -- & -- & 94.556 & 94.538 & 94.477 \\
	\multicolumn{1}{c}{} & \multicolumn{1}{c}{} & 2 & 7  & -- & \bf{96.647} & 96.626 & 96.600 & -- & -- & 83.283 & 83.282 & 96.646 & 96.627 & 96.569 \\
	\multicolumn{1}{c}{} & \multicolumn{1}{c}{4} & 2 & 20  & -- & 97.110 & 97.097 & 97.064 & -- & -- & 83.219 & 83.189 & \bf{97.111} & 97.096 & 97.051 \\
	\multicolumn{1}{c}{} & \multicolumn{1}{c}{} & 7 & 7  & -- & \bf{98.593} & 98.560 & 98.534 & -- & -- & 98.391 & 98.390 & 98.592 & 98.554 & 98.517 \\
	\multicolumn{1}{c}{} & \multicolumn{1}{c}{} & 7 & 20  & -- & \bf{98.986} & 98.957 & 98.931 & -- & -- & 98.844 & 98.843 & 98.984 & 98.948 & 98.918 \\
	\multicolumn{1}{c}{} & \multicolumn{1}{c}{} & 20 & 20  & -- & \bf{99.316} & 99.300 & 99.281 & -- & -- & 99.298 & 99.297 & 99.313 & 99.293 & 99.276 \\\hline
	\multicolumn{1}{c}{} & \multicolumn{1}{c}{} & 2 & 2  & -- & 89.366 & 89.548 & 89.691 & -- & -- & 85.221 & 85.189 & 89.377 & 89.602 & \bf{89.803} \\
	\multicolumn{1}{c}{} & \multicolumn{1}{c}{} & 2 & 7  & -- & 89.260 & 89.547 & 89.730 & -- & -- & 84.782 & 84.782 & 89.281 & 89.617 & \bf{89.802} \\
	\multicolumn{1}{c}{} & \multicolumn{1}{c}{2} & 2 & 20  & -- & 89.255 & 89.550 & 89.742 & -- & -- & 84.279 & 84.279 & 89.284 & 89.629 & \bf{89.912} \\
	\multicolumn{1}{c}{} & \multicolumn{1}{c}{} & 7 & 7  & 94.206 & 94.265 & 94.277 & 94.266 & 94.268 & 94.248 & 93.887 & 93.887 & 94.267 & \bf{94.277} & 94.239 \\
	\multicolumn{1}{c}{} & \multicolumn{1}{c}{} & 7 & 20  & 94.261 & 94.312 & 94.322 & 94.317 & 94.318 & 94.296 & 94.010 & 94.010 & 94.314 & \bf{94.323} & 94.295 \\
	\multicolumn{1}{c}{} & \multicolumn{1}{c}{} & 20 & 20  & 95.411 & 95.410 & 95.399 & 95.376 & 95.377 & \bf{95.412} & 95.332 & 95.332 & 95.409 & 95.394 & 95.341 \\
	\multicolumn{1}{c}{} & \multicolumn{1}{c}{} & 2 & 2  & -- & 93.417 & 93.710 & 93.810 & -- & -- & -- & -- & 93.442 & 93.754 & \bf{93.834} \\
	\multicolumn{1}{c}{} & \multicolumn{1}{c}{} & 2 & 7  & -- & 93.853 & 94.118 & 94.198 & -- & -- & 84.718 & 84.718 & 93.878 & 94.154 & \bf{94.215} \\
	\multicolumn{1}{c}{0.9} & \multicolumn{1}{c}{3} & 2 & 20  & -- & 93.898 & 94.208 & 94.300 & -- & -- & 84.788 & 84.788 & 93.930 & 94.244 & \bf{94.314} \\
	\multicolumn{1}{c}{} & \multicolumn{1}{c}{} & 7 & 7  & 97.704 & 97.714 & 97.692 & 97.664 & -- & \bf{97.716} & 97.468 & 97.468 & 97.714 & 97.688 & 97.636 \\
	\multicolumn{1}{c}{} & \multicolumn{1}{c}{} & 7 & 20  & 97.830 & \bf{97.847} & 97.833 & 97.810 & -- & -- & 97.610 & 97.610 & 97.847 & 97.830 & 97.774 \\
	\multicolumn{1}{c}{} & \multicolumn{1}{c}{} & 20 & 20  & \bf{98.582} & 98.568 & 98.545 & 98.513 & 98.535 & 98.575 & 98.522 & 98.522 & 98.567 & 98.536 & 98.480 \\
	\multicolumn{1}{c}{} & \multicolumn{1}{c}{} & 2 & 2  & -- & 95.965 & 96.019 & 95.985 & -- & -- & -- & -- & 95.972 & \bf{96.031} & 95.935 \\
	\multicolumn{1}{c}{} & \multicolumn{1}{c}{} & 2 & 7  & -- & 96.439 & 96.492 & 96.467 & -- & -- & 86.250 & 86.252 & 96.446 & \bf{96.500} & 96.441 \\
	\multicolumn{1}{c}{} & \multicolumn{1}{c}{4} & 2 & 20  & -- & 96.580 & 96.628 & 96.608 & -- & -- & 86.165 & 86.166 & 96.585 & \bf{96.640} & 96.581 \\
	\multicolumn{1}{c}{} & \multicolumn{1}{c}{} & 7 & 7  & -- & \bf{99.064} & 99.039 & 99.019 & -- & -- & 98.977 & 98.976 & 99.063 & 99.034 & 99.009 \\
	\multicolumn{1}{c}{} & \multicolumn{1}{c}{} & 7 & 20  & -- & \bf{99.183} & 99.162 & 99.133 & -- & -- & 99.107 & 99.106 & 99.181 & 99.152 & 99.119 \\
	\multicolumn{1}{c}{} & \multicolumn{1}{c}{} & 20 & 20  & -- & 99.527 & 99.515 & 99.509 & -- & -- & \bf{99.532} & 99.530 & 99.523 & 99.515 & 99.507 \\
\end{tabular}
}
\end{center}
\caption{
Average accuracy for each estimation method in Scenario 2 Case B with $n=400$ (\%). 
ML indicates the result based on the likelihood. 
The bold is the largest value for each setting. 
``--'' indicates that there were at least 100 cases in which the estimation algorithm did not converge or the estimate was not updated from the initial value. }
\label{tab:scenario2_caseb}
\end{table}

\section{Concluding remarks}
There are many previous studies on the robustness of regression models for binary responses, though fewer than for continuous responses. 
However, there are very few studies that discuss them from a bird's eye view. 
This study presents a framework to overview the properties of robust estimation based on divergence for generalized linear regression models for binary response. 
Our main results formulate the behavior of robustness for the linear model in terms of a link function and a divergence-based loss function when the linear predictors include contamination.
Methods based on $\beta$ and $\gamma$-divergence, often used in the context of robust statistics, are also robust in our analysis, as conventionally discussed. 
In particular, for $\gamma$-divergence, the loss function can be defined even when $\gamma$ is negative, and can be robust when $\gamma<-1$. 
It is also interesting to note that the case $\gamma=-2$ reduces to the Brier score (i.e., $L_2$-distance). 
It can also be shown that the loss function induced by $\gamma=-2$ is essentially the same as the one by an entropy induced by the harmonic mean.

While this study provided a new insight into robust estimation for binary regression models, it also highlighted several directions for future research. 
First, we found that the Brier score is an important index that links $\beta$ and $\gamma$-divergence. 
Based on this, a more comprehensive divergence based on a generalization of the Strictly Proper Scoring Rule is needed \cite{Merkle2013}. 
Second, the algorithm needs to be improved. 
The algorithm used in this study is based on \cite{Castilla2018}, but the loss function induced from the proposed divergence is in general not convex with respect to the parameters. 
To address this issue, it is necessary to consider introducing optimization methods, such as the minorize-minimization algorithm presented in \cite{Chi2014}. 
Lastly, generalization to multi-category responses is a challenge. 
Addressing these issues is expected to further extend the theoretical framework and practical applications of the proposed method. 
\bibliography{harmonic}
\bibliographystyle{plainnat}
\clearpage
\section*{Appendix}
\subsection*{A.1. Proof of Lemma \ref{lem:dist.boundedness}}
By the assumption, $\lim_{z\to\infty}e^{tz}\bar{G}(z)=0$ by l'H\^opital's rule. 
Hence we can write $\bar{G}(z)=exp(-h(z))$ where $h$ is a monotonically increasing function that grows faster than the linear order, and
\begin{align*}
L_1(c)
=
|z|h'(z)\exp(-(1-c)h(z)).
\end{align*}
When $1-c>0$, $L_1(c)$ decays exponentially because $h'$ grows smaller than $\exp(-(1-c)h(z))$. 
When $c=1$ then $L_1(c)=|z|h'(z)\to\infty$ as $z\to\infty$ because $h'$ tends to $\infty$.
For $L_2(c)$, the same argument can be applied.
\qed
\subsection*{A.2. Proof of Proposition \ref{prop:bias.boundedness}}
In this proof, we fix $y=1$ and $z'=0$ without loss of generality. 
First, consider Equation \eqref{eq:IF.linear.pred} for the loss function \eqref{eq:beta} and we have
\begin{align*}
|b_\beta(1,z,0)|
&\leq
2G(z)^\beta\frac{g(z)}{G(z)}|z|
 +2\bar{G}(z)^\beta\frac{g(z)}{\bar{G}(z)}|z|
=
2\frac{|z|g(z)}{G(z)^{1-\beta}}+2\frac{|z|g(z)}{\bar{G}(z)^{1-\beta}}.
\end{align*}
We shall evaluate the terms in the rightmost hand side of the above inequality. 
When $\beta\in(0,1)$, the first term converges to zero as $z\to\infty$ by the assumption of Proposition. 
The second term also converges to zero as $z\to\infty$ by Lemma \ref{lem:dist.boundedness}. 
When $\beta\geq1$, the two terms converges to zero by the assumption for $g$. 
The same argument is applied to the case $z\to-\infty$.

Next, consider the loss function \eqref{eq:gamma} and we have
\begin{align}\label{eq:bgamma}
|b_\gamma(1,z,0)|
&\leq
\frac{1}{\gamma+1}\left|G(z)_\gamma^{\frac{\gamma}{\gamma+1}}\frac{g_\gamma(z)}{G_\gamma(z)}z\right|
=
\frac{1}{\gamma+1}\left|\frac{zg_\gamma(z)}{G_\gamma(z)^{\frac{1}{\gamma+1}}}\right|,
\end{align}
where 
$
G_\gamma(z)=\dfrac{G(z)^{\gamma+1}}{G(z)^{\gamma+1}+\bar{G}(z)^{\gamma+1}}
$ and 
$
g_\gamma(z)=\dfrac{d}{dt}G_\gamma(z)
$. 
Obviously \eqref{eq:bgamma} diverges as $z\to-\infty$ when $\gamma=0$. 
We shall evaluate (un)boundedness of \ref{eq:bgamma} for four cases. 
\subsubsection*{case (i): $\gamma>0$}
It suffices to show that the density function $g_\gamma$ satisfies the assumption in Lemma \ref{lem:dist.boundedness}. 
By simple calculation we obtain  
\begin{align}\label{eq:g.gamma.density}
g_\gamma(z)
=
(\gamma+1)\frac{G_{\gamma+1}(z)\bar{G}_{\gamma+1}(z)}{G(z)\bar{G}(z)}g(z)
=
(\gamma+1)\frac{G(z)^{\gamma}\bar{G}(z)^{\gamma}}{(G(z)^{\gamma+1}+\bar{G}(z)^{\gamma+1})^2}g(z). 
\end{align}
Thus, $e^{tz}g_\gamma(z)\to0$ follows because of boundedness of the multiplier for $g$ in the rightmost hand side of Equation \eqref{eq:g.gamma.density}. 
\subsubsection*{case (ii): $\gamma\in(-1,0)$}
Reparamerize as $u=\gamma+1$ and $v=-\gamma$. 
Then we have $u,v>0$ and $u+v=1$. 
Now we have 
\begin{align*}
\frac{zg_\gamma(z)}{G_\gamma(z)^{\frac{1}{\gamma+1}}}
&\propto
z\left(\frac{G(z)^{v+1}+\bar{G}(z)^{v+1}}{G(z)^{v+1}}\right)^{\frac{1}{u}}
 \frac{G^{v+1}(z)\bar{G}^{v+1}(z)}{G(z)\bar{G}(z)}g(z)
\\
&=
\left(G(z)^{v+1}+\bar{G}(z)^{v+1}\right)^{\frac{1}{u}}
 \left(G(z)^{\frac{v}{u}}\bar{G}^{\frac{v}{u}}(z)\right)^{\frac{1}{u}}
  \frac{zg(z)}{G(z)^{\frac{1+v}{u}}}.
\end{align*}
Hence, there exists a positive value $M$ such that 
$
\left|\dfrac{zg_\gamma(z)}{G_\gamma(z)^{\frac{1}{\gamma+1}}}\right|
\leq M \left|\dfrac{zg(z)}{G(z)^{\frac{1+v}{u}}}\right|
\to0
$
as $z\to-\infty$. 
\subsubsection*{case (iii): $\gamma=-1$}
By the loss function \eqref{eq:geometric}, we have
\begin{align*}
\lim_{\gamma\to-1}|b_{\gamma}(1,z,0)|
=
\frac{1}{2}\sqrt{\frac{G(z)}{\bar{G}(z)}}
 \left| \frac{zg(z)}{\bar{G}(z)}-\frac{zg(z)}{G(z)}
  \right|
\leq
\frac{1}{2\sqrt{G(z)\bar{G}(z)}}
 \left| \frac{zg(z)}{\bar{G}(z)} \right|
  +\left|\frac{zg(z)}{2\sqrt{G(z)\bar{G}(z)}}\right|.
\end{align*}
Since $\dfrac{|z|g(z)}{\bar{G}(z)}$ diverges by the assumption and as $z\to\infty$ and $\dfrac{1}{\sqrt{G(z)\bar{G}(z)}}$  is unbounded, 
the upper bound of the above inequality is unbounded. 
\subsubsection*{case (iv): $\gamma<-1$}
There exists $u>0$ such that $\gamma+1=-u<0$, hence we have
\begin{align*}
\left|\frac{zg_\gamma(z)}{G_\gamma(z)^{\frac{1}{\gamma+1}}}\right|
=
\left|\frac{zg_{-u-1}(z)}{G_{-u-1}(z)^{-\frac{1}{u}}}\right|
=
\left|u\left(\frac{G(z)^{u}}{G(z)^{u}+\bar{G}(z)^{u}}\right)
 \left(\frac{\bar{G}(z)^{u}}{G(z)^{u}+\bar{G}(z)^{u}}\right)
  \left(\frac{1}{G(z)^{u}+\bar{G}(z)^{u}}\right)^{\frac{1}{u}}
  \frac{zg(z)}{G(z)}\right|.
\end{align*}
Thus, the above equation converges to zero by Lemma \ref{lem:dist.boundedness}. 
\qed

\end{document}